\documentclass[12pt]{article}
\usepackage{epsfig}
\usepackage{axodraw}

\begin{document}
\title{Semileptonic Decay
of  $B$ and $D\to K^*_0(1430) \bar{\ell}\nu$ From QCD Sum Rule}
\author{ Mao-Zhi Yang
 \thanks{Email address: yangmz@mail.ihep.ac.cn (M.Z.Yang)}\\
{\small  CCAST (World Laboratory), P.O. Box 8730, Beijing 100080, China}\\
{\small and }\\
{\small  Institute of High Energy Physics, Chinese
Academy of Sciences,}\\
{\small P.O. Box 918(4), Beijing 100049,China}\footnote{Mailing
address} }
\date{\empty}
\maketitle

\begin{abstract}
We calculate $B_{(s)}$ and $D_{(s)}$ to $K^*_0(1430)$ transition
form factors, and study semileptonic decays of $B_{(s)}$ and
$D_{(s)}\to K_0^*(1430) \bar{\ell}\nu$ based on QCD sum rule.
Measuring these semileptonic decays with high statistics will give
valuable information on the nature of light scalar mesons.
\end{abstract}

\hspace{1cm} \small{PACS numbers: 12.38.Lg, 13.20.Fc,13.20.He}

\vspace{1cm}

Semileptonic decays of $B$ and $D$ mesons are important for
studying quark flavor mixing and extracting the
Cabibbo-Kobayashi-Maskawa (CKM) matrix element, because strong
interactions involved in these processes are simpler than that of
hadronic decays. The strong binding effects can be parameterized
as transition form factors, which can be calculated by Lattice QCD
\cite{LQCD}, QCD sum rule \cite{SVZ,sum,BBD,dly}, light-cone sum
rule \cite{LCSR}, or by quark model \cite{QuarkModel}, light-front
approach \cite{LightFront}, and recently by large-energy and
heavy-quark-effective theory \cite{fk}. Experimental data on
semileptonic decay of $B$ and $D$ mesons can be used to test these
theoretical method of treating nonperturbative dynamics.
Transition form factors inducing semileptonic decays not only
depend on dynamics of strong interactions between quarks in the
initial and final hadrons, but also on the structure of the
hadrons involved in the semileptonic decays.

The structures of scalar mesons are long-standing problems in
particle physics. A large amount of scalar mesons have been found
in experiment \cite{PDG}. They include
 $\sigma$ [or $f_0(600)$],
$f_0(980)$, $f_0(1370)$, $f_0(1500)$, $f_0(1710)$, $a_0(980)$,
$a_0(1450)$, $\kappa$, $K_0^*(1430)$, etc. \cite{span}. At least
they can be divided into two flavor nonets, one below or near $1
\mathrm{GeV}$, the other above $1 \mathrm{GeV}$. The structure of
scalar mesons is still not well established theoretically
\cite{scalar}. In the literature, many suggestions are discussed
such as $q\bar{q}$, $q\bar{q}q\bar{q}$ and meson-meson bound
states. Among the controversy, almost every model on scalar states
agrees that $K_0^*(1430)$ is dominated by $s\bar{u}$ or $s\bar{d}$
state. Recent analysis from QCD sum rule agrees with this picture
\cite{plb619}. The result of QCD sum rule favors that
$K_0^*(1430)$ is mainly $s\bar{q}$ bound state. The predicted mass
of $s\bar{q}$ $0^+$ scalar bound state is consistent with the mass
of $K_0^*(1430)$.

To investigate the structure of scalar meson, a large amount of
experimental data and theoretical studies are necessary. scalar
mesons are produced in $\pi N$ scattering, $p\bar{p}$
annihilation, decays of heavy flavor mesons, etc.. Recently,
hadronic $D$ decays involving scalar mesons are studied in
generalized factorization approach \cite{cheng1}. Hadronic decay
of $D$ meson is complicated, it suffers from final-state
interactions which are difficult to control theoretically.
Compared with hadronic decays, semleptonic decays are more
simpler. They are free from final-state interactions. All
nonperturbative binding effects are parameterized in the
transition form factors.

In this work, we shall study semileptonic decays of $B_{(s)}$ and
$D_{(s)}$ meson involving $K_0^*(1430)$. In Ref.\cite{ss},
semileptonic decays of $D$ meson into $\sigma$, $\kappa$ are
studied in QCD sum rule, where the scalars $\sigma$ and $\kappa$
are treated as $q\bar{q}$ bound states, which is different from
the picture that scalars below 1 GeV are predominantly multiquark
states. According to our previous study in QCD sum rule,
$K_0^*(1430)$ is dominantly the ground state of $s\bar{q}$ scalar
channel. We treat $K_0^*(1430)$ as $s\bar{q}$ bound state in this
work. The transition form factors of $B_{(s)}$ and $D_{(s)}\to
K_0^*(1430)$ are calculated in QCD sum rule. Then they are used to
study semileptonic $B$ and $D$ decays. The semileptonic decay
modes of $B_{(s)}$ and $D_{(s)}$ mesons involving $K_0^*(1430)$
include $B_s^0\to K_0^*(1430)^-\bar{\ell}\nu$, $D^0\to
K_0^*(1430)^-\bar{\ell}\nu$, $D^+\to
\bar{K}_0^*(1430)^0\bar{\ell}\nu$, and $D_s^+\to
K_0^*(1430)^0\bar{\ell}\nu$, where $\ell$ is the lepton $e$ or
$\mu$. Compared to the large mass of $B$ and $D$ meson, the mass
of the lepton $e$ or $\mu$ is dropped.

\begin{figure}[h]
\begin{center}
\begin{tabular}{cc}
\scalebox{0.5}{
\begin{picture}(265,155) (120,-60)
\SetWidth{0.5} \Vertex(195,-15){2.83}
\ZigZag(195,-15)(240,30){7.5}{3} \ArrowLine(285,75)(240,30)
\ArrowLine(240,30)(300,30) \Vertex(240,30){2.83}
\ArrowLine(315,-15)(195,-15) \ArrowLine(195,-15)(120,-15)
\ArrowLine(120,-60)(195,-60) \ArrowLine(195,-60)(315,-60)
\put(290,80){{\LARGE $\bar{\ell}$}} \put(310,30){{\LARGE $\nu$}}
\put(110,-15){{\LARGE $\bar{b}$}} \put(320,-15){{\LARGE
$\bar{u}$}} \put(110,-60){{\LARGE $s$}}\put(320,-60){{\LARGE $s$}}
\put(210,-90){{\LARGE $(a)$}}
\end{picture}}

\scalebox{0.5}{
\begin{picture}(265,155) (120,-60)
\SetWidth{0.5} \Vertex(195,-15){2.83}
\ZigZag(195,-15)(240,30){7.5}{3} \ArrowLine(285,75)(240,30)
\ArrowLine(240,30)(300,30) \Vertex(240,30){2.83}
\ArrowLine(195,-15)(315,-15) \ArrowLine(120,-15)(195,-15)
\ArrowLine(195,-60)(120,-60) \ArrowLine(315,-60)(195,-60)
\put(290,80){{\LARGE $\bar{\ell}$}} \put(310,30){{\LARGE $\nu$}}
\put(110,-15){{\LARGE $c$}} \put(320,-15){{\LARGE $s$}}
\put(110,-60){{\LARGE $\bar{u}$}}\put(320,-60){{\LARGE $\bar{u}$}}
\put(210,-90){{\LARGE $(b)$}}
\end{picture}}\\

\scalebox{0.5}{
\begin{picture}(265,200) (120,-60)
\SetWidth{0.5} \Vertex(195,-15){2.83}
\ZigZag(195,-15)(240,30){7.5}{3} \ArrowLine(285,75)(240,30)
\ArrowLine(240,30)(300,30) \Vertex(240,30){2.83}
\ArrowLine(195,-15)(315,-15) \ArrowLine(120,-15)(195,-15)
\ArrowLine(195,-60)(120,-60) \ArrowLine(315,-60)(195,-60)
\put(290,80){{\LARGE $\bar{\ell}$}} \put(310,30){{\LARGE $\nu$}}
\put(110,-15){{\LARGE $c$}} \put(320,-15){{\LARGE $s$}}
\put(110,-60){{\LARGE $\bar{d}$}}\put(320,-60){{\LARGE $\bar{d}$}}
\put(210,-90){{\LARGE $(c)$}}
\end{picture}}

\scalebox{0.5}{
\begin{picture}(265,200) (120,-60)
\SetWidth{0.5} \Vertex(195,-15){2.83}
\ZigZag(195,-15)(240,30){7.5}{3} \ArrowLine(285,75)(240,30)
\ArrowLine(240,30)(300,30) \Vertex(240,30){2.83}
\ArrowLine(195,-15)(315,-15) \ArrowLine(120,-15)(195,-15)
\ArrowLine(195,-60)(120,-60) \ArrowLine(315,-60)(195,-60)
\put(290,80){{\LARGE $\bar{\ell}$}} \put(310,30){{\LARGE $\nu$}}
\put(110,-15){{\LARGE $c$}} \put(320,-15){{\LARGE $d$}}
\put(110,-60){{\LARGE $\bar{s}$}}\put(320,-60){{\LARGE $\bar{s}$}}
\put(210,-90){{\LARGE $(d)$}}
\end{picture}}
\end{tabular}
\end{center}
\caption{Semileptonic decays of $B_{(s)}$ and $D_{(s)}$ involving
$K_0^*(1430)$. (a) for $B_s^0\to K_0^*(1430)^-\bar{\ell}\nu$, (b)
for $D^0\to K_0^*(1430)^-\bar{\ell}\nu$, (c) for $D^+\to
\bar{K}_0^*(1430)^0\bar{\ell}\nu$, and (d) for $D_s^+\to
K_0^*(1430)^0\bar{\ell}\nu$ } \label{fig1}
\end{figure}
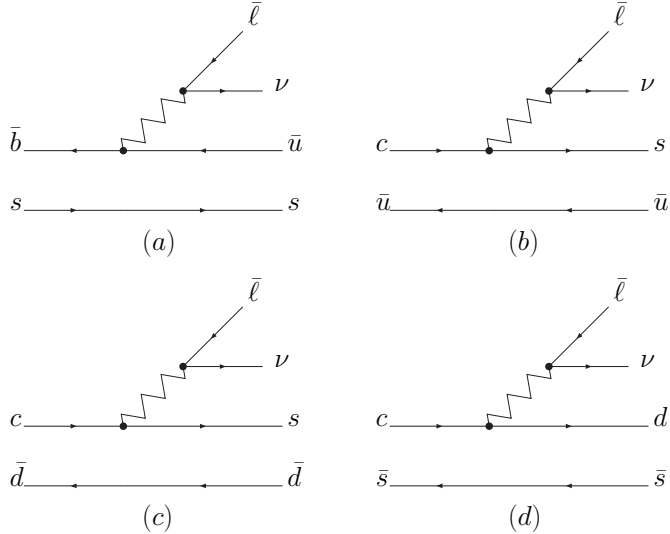

The Feynman diagrams for the semileptonic $B_{(s)}$ and $D_{(s)}$
decays into $K_0^*(1430)\bar{\ell}\nu$ are shown in
Fig.(\ref{fig1}). In the limit of flavor SU(3) symmetry, the
amplitudes of the three decay modes $D^0\to
K_0^*(1430)^-\bar{\ell}\nu$,
$D^+\to\bar{K}_0^*(1430)^0\bar{\ell}\nu$ and $D_s^+\to
K_0^*(1430)^0\bar{\ell}\nu$ should be the same. However, flavor
SU(3) symmetry is not an exact symmetry, because the mass of $s$
quark is much larger than that of $u$ and $d$ quarks. To
investigate SU(3) symmetry breaking effect, the mass of $s$ quark
$m_s$ is kept in our calculation.

The amplitude of $B$ and $D$ decaying into
$K_0^*(1430)\bar{\ell}\nu$ is
\begin{equation}\label{1}
 A=\frac{G_F}{\sqrt{2}}V_{Qq'}^*\bar{\nu}\gamma_\mu (1-\gamma_5)\ell
 \langle K_0^*(1430)|\bar{q}'\gamma^\mu (1-\gamma_5)Q|H \rangle
 \end{equation}
where $G_F$ is Fermi constant, $H$ is the heavy flavor meson $B$
or $D$, $Q$ is $b$ or $c$ quark, $q'$ can be light quark $u$, $d$,
$s$, and $V_{Qq'}$ is the relevant CKM matrix element. Strong
interactions are contained in the hadronic matrix element $\langle
K_0^*(1430)|\bar{q}'\gamma^\mu (1-\gamma_5)Q|H\rangle$. By
analyzing the parity property of the hadrons and the current, we
can know that the vector current does not contribute to the
pseudoscalar-scalar hadronic matrix element
\begin{equation}\label{2}
 \langle K_0^*(1430)|\bar{q}'\gamma^\mu Q|H\rangle =0,
 \end{equation}
therefore, only the axial current contributes. From the Lorentz
invariance, the hadronic matrix element of axial current can be
decomposed into
\begin{equation}\label{3}
 \langle K_0^*(1430)|\bar{q}'\gamma^\mu \gamma_5 Q|H\rangle
 =-i[ (p_1+p_2)^\mu F_+(q^2) +(p_1-p_2)^\mu F_-(q^2)],
\end{equation}
where the parameters $F_+(q^2)$ and $ F_-(q^2)$ are the transition
form factors, which only depend on the momentum transfer squared.
The momenta $p_1$ and $p_2$ are the momenta of the initial and
final state mesons, respectively, and $q=p_1-p_2$.

Next we shall calculate the transition form factors in QCD sum
rule method. In the limit of the lepton mass $m_\ell\sim 0$, the
form factor $F_-$ does not contribute to the semileptonic decays
of $B$ and $D$ mesons. So we shall not consider $F_-$ in this
work.

In the standard procedure of QCD sum rule for calculating
transition form factors, all hadrons involved in the transition
process are interpolated by currents with the appropriate quantum
numbers relevant to the hadrons. The heavy flavor meson $H$ ($B$
or $D$) is interpolated by pseudoscalar current
$J_H(x)=\bar{Q}(x)i\gamma_5 q'(x)$, summation over the spinor and
color indices being understood in this interpolating current. The
final scalar meson $K_0^*(1430)$ is interpolated by the scalar
current $J_K(x)=\bar{q}(x)s(x)$, here $q$ is up or down quark.
Then the decay rate is obtained from the vacuum expectation value
of time ordered product of the interpolating fields $J_H(x)$,
$J_K(x)$ and the weak axial current
$J_{\mu}=\bar{q}'\gamma_\mu\gamma_5 Q$. The vector part of the
weak current is not presented here because its contribution to $B$
or $D\to K_0^*(1430)$ transition is zero. A three-point
correlation function can be constructed from the vacuum
expectation value of time ordered product of interpolating fields
and axial weak current
\begin{equation}
\Pi_{\mu}(p_1^2,p_2^2,q^2)=i^2\int d^4 x d^4 y e^{ip_2\cdot
x-ip_1\cdot y}
 \langle 0|T\{J_K(x) J_{\mu}(0)J_H(y)\} |0\rangle .
\label{correlator}
\end{equation}
Phenomenologically, one can insert a full set of intermediate
sates into the time ordered product and obtain the double
dispersion relation, then the correlation function is expressed as
a sum of the contributions of the lowest lying and excited states
\begin{eqnarray}
\Pi_{\mu}(p_1^2,p_2^2,q^2)&=&\frac{\langle
0|\bar{q}'_2q'_1|K_0^*(1430)\rangle \langle
K_0^*(1430)|\bar{q}'_1\gamma_\mu \gamma_5 Q|H\rangle \langle
H|\bar{Q}i\gamma_5
q'_2|0\rangle}{(m_H^2-p_1^2)(m_{K_0^*}^2-p_2^2)}\nonumber\\
&&+\mathrm{excited~ states},
\end{eqnarray}
where $q'_1$ and $q'_2$ can be $u$, $d$, $s$ quarks, but they do
not take the same flavor at the same time and there must be one to
be $s$ quark in $q'_1$ and $q'_2$. By introducing $\langle
0|\bar{q}s|K_0^*(1430)\rangle =f_{K_0^*}m_{K_0^*}$ and $\langle
H|\bar{Q}i\gamma_5 q'|0\rangle =\frac{m_H^2}{m_Q+m_{q'}}f_H$,
where $f_{K_0^*}$ and $f_H$ are decay constants, the correlation
function is obtained as
\begin{eqnarray}\label{ch}
\Pi_{\mu}(p_1^2,p_2^2,q^2)&=&\frac{m_H^2}{m_Q+m_{q'_2}}\frac{f_H
f_{K_0^*}m_{K_0^*}} {(m_H^2-p_1^2)(m_{K_0^*}^2-p_2^2)}\langle
K_0^*(1430)|\bar{q}'_1\gamma_\mu \gamma_5 Q|H\rangle \nonumber\\
&&+\mathrm{excited~ states}
\end{eqnarray}
Therefore the correlation function is related to the transition
matrix element by eq.(\ref{ch}). If the correlation function
$\Pi_{\mu}(p_1^2,p_2^2,q^2)$ can be calculated in QCD reliably,
then the transition form factors can be extracted through
eq.(\ref{ch}). In QCD, the three-point correlation function can
indeed be evaluated by operator-product-expansion (OPE) method in
the deep Euclidean region
\begin{equation}
p_1^2\ll (m_Q+m_{q'_2})^2,~~p_2^2\ll (m_s+m_q)^2.
\end{equation}
Although it is in the deep Euclidean region of $p_1^2$ and $p_2^2$
that the correlation function can be calculated, eq.(\ref{ch})
clearly shows that the value of $\Pi_{\mu}(p_1^2,p_2^2,q^2)$ can
still be used to extract the transition form factors which is
typically a Minkowskian quantity, because the valuables $p_1^2$
and $p_2^2$ are clearly separated from the transition matrix
element $\langle K_0^*(1430)|\bar{q}'_1\gamma_\mu \gamma_5
Q|H\rangle$ in the left hand of eq.(\ref{ch}). Next we shall
evaluate the correlation function in QCD.

 The time-ordered current operators in the three-point
correlation function in eq.(\ref{correlator}) can be expanded in
terms of a series of local operators with increasing dimensions,
\begin{eqnarray}
&&i^2\int d^4 x d^4 y e^{ip_2\cdot x-ip_1\cdot y}
 T\{J_K(x) J_{\mu} (0)J_H(y)\}  \nonumber\\
 &=&C_{0\mu} I +C_{3\mu} \bar{\Psi}\Psi
    +C_{4\mu} G^a_{\alpha\beta}G^{a\alpha\beta}
    +C_{5\mu} \bar{\Psi}\sigma_{\alpha\beta}T^a G^{a\alpha\beta}\Psi
    \nonumber\\
  &~+&C_{6\mu}
 \bar{\Psi}\Gamma \Psi \bar{\Psi}\Gamma^{\prime}\Psi+\cdots,
 \label{opef}
 \end{eqnarray}
where $C_{i\mu}$'s are Wilson coefficients, $I$ is the unit
operator, $\bar{\Psi}\Psi$ is the local Fermion field operator of
light quarks, $G^a_{\alpha\beta}$ is gluon strength tensor,
$\Gamma$ and $\Gamma^{\prime}$ are the matrices appearing in the
procedure of calculating the Wilson coefficients. Considering the
vacuum expectation value of the OPE of
the interpolating field
operator, we get the correlation function in terms of Wilson
coefficients and condensates of local operators,
\begin{eqnarray}
\Pi_{\mu}(p_1^2,p_2^2,q^2)&=&i^2\int d^4x d^4 y e^{ip_2\cdot
x-ip_1\cdot y}
 \langle 0|T\{J_K(x) J_{\mu} (0)J_H(y)\}|0\rangle \nonumber\\
 &=&C_{0\mu} +C_{3\mu} \langle 0|\bar{\Psi}\Psi|0\rangle
    +C_{4\mu} \langle 0|G^a_{\alpha\beta}G^{a\alpha\beta}|0\rangle
    +C_{5\mu} \langle 0|\bar{\Psi}\sigma_{\alpha\beta}T^a G^{a\alpha\beta}\Psi|0\rangle
    \nonumber\\
  &~+&C_{6\mu}\langle 0|
 \bar{\Psi}\Gamma \Psi \bar{\Psi}\Gamma^{\prime}\Psi|0\rangle +\cdots,
 \label{conden}
\end{eqnarray}
The Wilson coefficients $C_i$'s depend only on $p_1$ and $p_2$,
according to the Lorentz structure of the correlation function,
the result can be re-expressed by two parts in the ``theoretical
expression"
\begin{equation}\label{ct}
\Pi_{\mu}(p_1^2,p_2^2,q^2)=-i f_+(p_1+p_2)_\mu-i f_-(p_1-p_2)_\mu,
\end{equation}
The coefficients $f_{\pm}$ above collect all the contributions of
perturbative and condensate terms
\begin{equation}
f_{\pm}=f_{\pm}^{pert}+f_{\pm}^{(3)}+f_{\pm}^{(4)}+f_{\pm}^{(5)}
+f_{\pm}^{(6)}+\cdots
\end{equation}
where $f_{\pm}^{pert}$ is the perturbative contribution of the
unit operator, and $f_i^{(3)}$, $\cdots$, $f_i^{(6)}$ are
contributions of condensates of dimension 3, 4, 5, 6, $\cdots$
operators in the OPE. The perturbative contribution and gluon
condensate contribution can be written in the from of dispersion
integration,
\begin{eqnarray}
 f_{\pm}^{pert}&=&\int d s_1
 d s_2\frac{\rho^{pert}_{\pm}(s_1,s_2,q^2)}{(s_1-p_1^2)(s_2-p_2^2)},
 \nonumber  \\
f_{\pm}^{(4)}&=&\int d s_1
 d s_2\frac{\rho^{(4)}_{\pm}(s_1,s_2,q^2)}{(s_1-p_1^2)(s_2-p_2^2)}.
 \nonumber  \nonumber
 \end{eqnarray}
We can approximate the contribution of excited states as
integrations over some thresholds $s_1^0$ and $s_2^0$ in the above
equations, which is the assumption of quark-hadron duality
\cite{SVZ,175}. Then equate the ``phenomenological" and
``theoretical" expressions of the correlation function in
Eq.~(\ref{ch}) and (\ref{ct}), we can get an equation for the form
factors. But such equation may highly depend on the approximation
for the contribution of excited states and the condensate of
higher dimensional operators in OPE. To improve such equation, one
can make Borel transformation over $p_1^2$ and $p_2^2$ in both
sides, which can suppress the contributions of excited states and
condensate of higher dimensional operators. The definition of
Borel transformation to any function $f(p^2)$ is
$$\hat{B}_{\left|\frac{}{}\right.p^2,M^2}f(p^2)=\lim_{\small\begin{array}{ll}& n\to\infty \\
    & p^2\to -\infty  \\&-p^2/n=
    M^2  \end{array} } \frac{(-p^2)^n}{(n-1)!}\frac{\partial ^n}{\partial (p^2)^n}
    f(p^2).$$
Some examples of Borel transformation is given in the following,
\begin{eqnarray*}
&&\hat{B}_{\left|\frac{}{}\right.p^2,M^2}\frac{1}{(s-p^2)^k}=
\frac{1}{(k-1)!}\frac{1}{(M^2)^k}e^{-s/M^2},\\
&&\hat{B}_{\left|\frac{}{}\right.p^2,M^2}(p^2)^k=0,~~~ \mbox{for
any}~~ k\ge 0.
\end{eqnarray*}

Equating the two expressions of the correlation function,
subtracting the contribution of excited states, and performing
Borel transformation in both variables $p_1^2$ and $p_2^2$, we
finally obtain the sum rules for the form factor
\begin{equation}\label{formfactor}
F_+(q^2)=\frac{m_Q+m_{q'_2}}{f_{K_0^*}m_{K_0^*}m_H^2f_H}M_1^2M_2^2
   e^{m_H^2/M_1^2}e^{m_{K_0^*}^2/M_2^2}\hat{B}f_+,
 \end{equation}
where $\hat{B} f_{+}$ denotes Borel transforming $f_{+}$ in both
variables $p_1^2$ and $p_2^2$, $M_1$ and $M_2$ are Borel
parameters. After subtracting the contribution of the excited
states, now the dispersion integration for perturbative and gluon
condensate contribution should be performed under the threshold,
\begin{eqnarray}
 f_{+}^{pert}&=&\int^{s_1^0} d s_1
 \int^{s_2^0} d s_2\frac{\rho^{pert}_{+}(s_1,s_2,q^2)}{(s_1-p_1^2)(s_2-p_2^2)},
 \nonumber  \\
f_{+}^{(4)}&=&\int^{s_1^0} d s_1
 \int^{s_2^0}d s_2\frac{\rho^{(4)}_{+}(s_1,s_2,q^2)}{(s_1-p_1^2)(s_2-p_2^2)}.
 \nonumber  \nonumber
 \end{eqnarray}

\begin{figure}[h]
\vspace{1cm}
\begin{center}\scalebox{0.7}{
\begin{picture}(300,200)(-15,-25.98)
\DashLine(-11.25,100.515)(0,120){3} \Line(0,120)(30,171.96)
\Line(0,120)(30,171.96)\Line(30,171.96)(60,120)\Line(60,120)(0,120)
\DashLine(60,120)(71.25,100.515){3}\Photon(30,171.96)(30,190.71){1}{4}

\GlueArc(0,120)(30,40,60){2}{2}\GlueArc(0,120)(30,0,20){2}{2}
\put(19.88,137,25){$\times$}\put(23.5,129){$\times$}

\DashLine(88.75,100.515)(100,120){3} \Line(100,120)(130,171.96)
\Line(100,120)(130,171.96)\Line(130,171.96)(160,120)\Line(160,120)(100,120)
\DashLine(160,120)(171.25,100.515){3}\Photon(130,171.96)(130,190.71){1}{4}

\GlueArc(130,171.96)(30,240,260){2}{2}\GlueArc(130,171.96)(30,280,300){2}{2}
\put(120.4,138.8){$\times$}\put(131.3,138.8){$\times$}

\DashLine(188.75,100.515)(200,120){3} \Line(200,120)(230,171.96)
\Line(200,120)(230,171.96)\Line(230,171.96)(260,120)\Line(260,120)(200,120)
\DashLine(260,120)(271.25,100.515){3}\Photon(230,171.96)(230,190.71){1}{4}

\GlueArc(260,120)(30,160,180){2}{2}\GlueArc(260,120)(30,120,140){2}{2}
\put(232.3,136){$\times$}\put(228.125,128){$\times$}

\DashLine(-11.25,-19.485)(0,0){3} \Line(0,0)(30,51.96)
\Line(0,0)(30,51.96)\Line(30,51.96)(60,0)\Line(60,0)(0,0)
\DashLine(60,0)(71.25,-19.485){3}\Photon(30,51.96)(30,70.71){1}{4}

\Gluon(11.25,19.485)(-1.74,26.985){2}{2}\Gluon(18.75,32.475)(5.76,39.975){2}{2}

\put(-5.625,23.5){$\times$}\put(1.65,36.5){$\times$}
 \DashLine(88.75,-19.485)(100,0){3} \Line(100,0)(130,51.96)
\Line(100,0)(130,51.96)\Line(130,51.96)(160,0)\Line(160,0)(100,0)
\DashLine(160,0)(171.25,-19.485){3}\Photon(130,51.96)(130,70.71){1}{4}

\Gluon(122.5,0)(122.5,-15){2}{2}\Gluon(137.5,0)(137.5,-15){2}{2}

\put(119.875,-16.5){$\times$}\put(135.25,-16.5){$\times$}

\DashLine(188.75,-19.485)(200,0){3} \Line(200,0)(230,51.96)
\Line(200,0)(230,51.96)\Line(230,51.96)(260,0)\Line(260,0)(200,0)
\DashLine(260,0)(271.25,-19.485){3}\Photon(230,51.96)(230,70.71){1}{4}

\Gluon(248.75,19.485)(261.74,26.985){2}{2}\Gluon(241.25,32.475)(254.24,39.975){2}{2}

\put(256.5,24.75){$\times$}\put(250,37.5){$\times$}

\put(23,90){(a)}
 \put(125,90){(b)}
 \put(225,90){(c)}

 \put(23,-35){(d)}
 \put(125,-35){(e)}
 \put(225,-35){(f)}
\end{picture}}
\end{center}
\vspace{0.5cm} \caption{\small Diagrams for contributions of
bi-gluon operator.} \label{bigluon}
\end{figure}
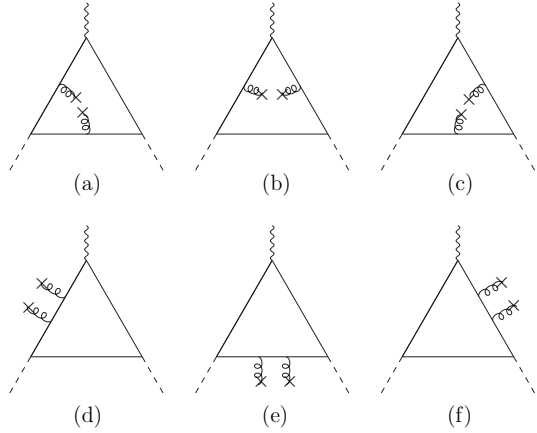

In this work we calculate the Wilson coefficients in OPE up to the
contribution of the operator of dimension 6. The diagrams we
considered here are similar to our previous work in
Ref.\cite{dly}. We find again that the contributions of the
diagrams for the gluon-gluon condensate in fig.\ref{bigluon}
exactly cancel in the sum in the case of pseudoscalar to scalar
transition. Therefore, the gluon-gluon condensate does not
contribute to $B$ and $D\to K_0^*(1430)$ transition form factors.

The results of Borel transformed coefficient $\hat{B}f_+$ in
eq.(\ref{formfactor}) are given in the Appendix.

Since we do not take into account radiative corrections in QCD, we
choose the parameters at a fixed renormalization scale of about 1
GeV. The values of the condensates are taken as \cite{SVZ,
Narison},
\begin{eqnarray}
&\langle \bar{q}q\rangle =-(0.24\pm 0.01 ~\mbox{GeV})^3, ~~~~
\langle \bar{s}s\rangle =m_0^2 \langle \bar{q}q\rangle,
\nonumber\\[4mm]
 & g\langle \bar{\Psi}\sigma TG\Psi \rangle =m_0^2 \langle
\bar{\Psi}\Psi \rangle, ~~~~\alpha_s\langle \bar{\Psi}\Psi\rangle
^2= 6.0\times10^{-5}~\mbox{GeV}^6 ,\\[4mm]
& m_0^2=0.8\pm 0.2~\mbox{GeV}^2. \nonumber
\end{eqnarray}

The quark masses are taken to be $m_b=4.7\pm 0.1~\mbox{GeV}$,
$m_c=1.3\pm 0.1 ~\mbox{GeV}$, $m_s=150\pm 10~\mbox{MeV}$ and
$m_u\sim m_d\sim 0$. The decay constants of $B_{(s)}$, $D_{(s)}$
and $K_0^*(1430)$ are taken from the two-point QCD sum rule:
$f_D=200\pm 20~\mbox{MeV}$, $f_B=180\pm 30 ~\mbox{MeV}$
\cite{175,443}, and from the ratios $f_{D_s}/f_D=1.19\pm 0.08$,
$f_{B_s}/f_B=1.16\pm 0.09$ \cite{175}, we can obtain
$f_{D_s}=238\pm 29~\mbox{MeV}$, $f_{B_s}=209\pm 38 ~\mbox{MeV}$.
For the decay constant of $K_0^*(1430)$, we get $f_{K_0^*}=427\pm
85~\mbox{MeV}$ from two-point sum rule \cite{plb619}.

The threshold parameters $s_1^0$ and $s_2^0$ are also determined
from the two-point sum rule. The parameter $s_1^0$ is for the
threshold of $B$ and $D$ meson, they are taken to be $s_B^0=35\pm
2~\mbox{GeV}^2$, $s_D^0=6\pm 1~\mbox{GeV}^2$ \cite{443}. $s_2^0$
is for the threshold of $K_0^*(1430)$, its value is $s_2^0=4.4\pm
0.4~\mbox{GeV}^2$ \cite{plb619}.

\begin{figure}[h]
\begin{center}
\epsfig{file=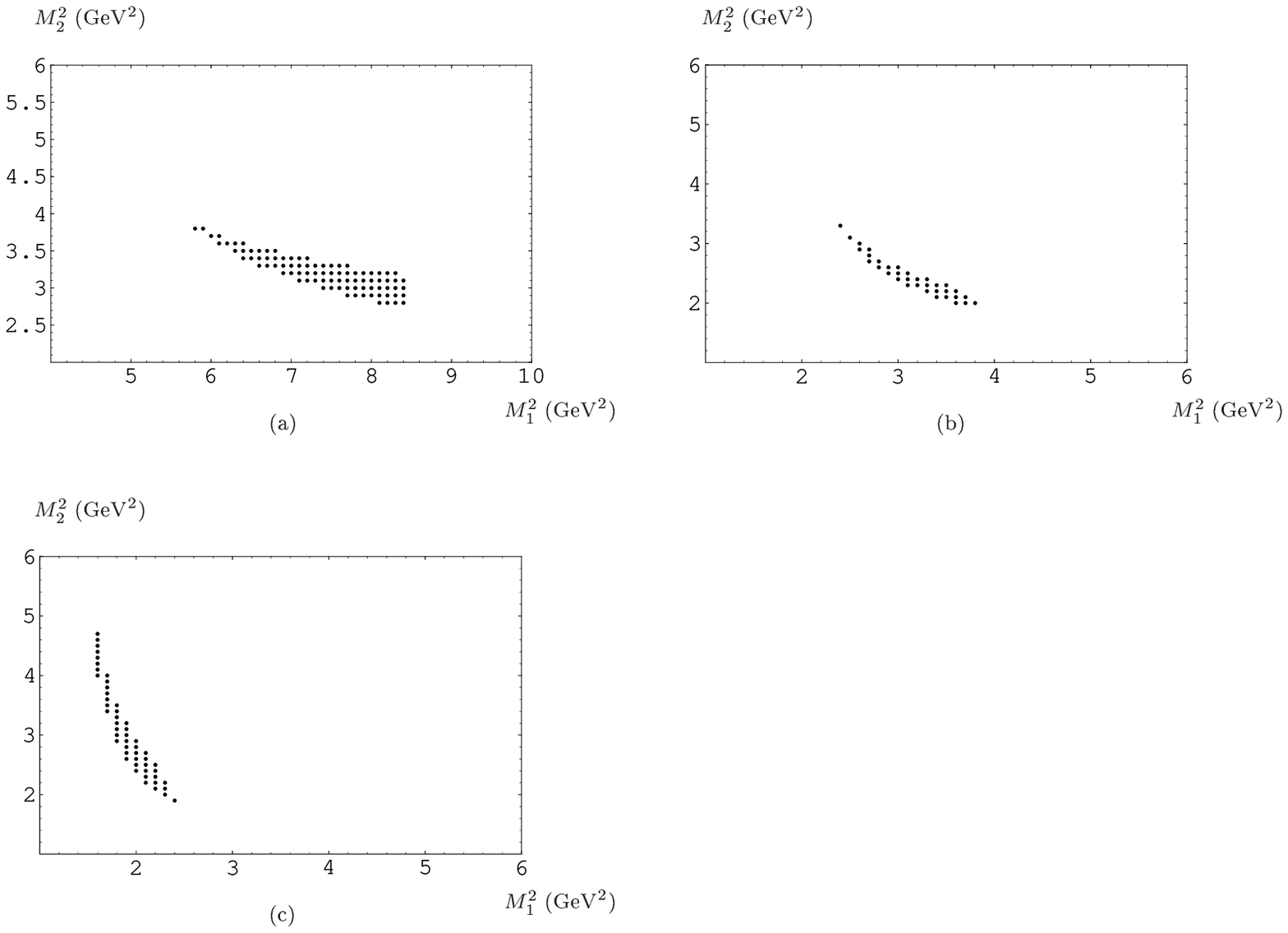, width=12cm,height=9cm}
\end{center}
\caption{\small Selected regions of $M_1^2$ and $M_2^2$: (a) for
$F_+^{B_sK_0^*}$; (b) for $F_+^{DK_0^*}$; (c) for
$F_+^{D_sK_0^*}$.} \label{region}
\end{figure}

The Borel parameters $M_1$ and $M_2$ are not physical parameters.
The physical result should not depend on them if the operator
product expansion can be calculated up to infinite order. However,
OPE has to be truncated to some finite orders in practice.
Therefore, Borel parameters have to be selected in some ``windows"
to get the best stability of the physical results. We choose $M_1$
and $M_2$ in the region where 1) the contribution of the excited
states is effectively suppressed, which can ensure that the sum
rule does not sensitively depend on the approximation for the
excited states, and 2) the contribution of the condensates should
not be too large, which can ensure that the contribution of the
higher dimensional operators is small and the truncated OPE is
effective. The contribution of the excited sates to the
three-point correlation function is in the form
$e^{-m^2_{excited}/M^2_{1,2}}$, where $m_{excited}$ denotes the
mass of the excited state, and the series in OPE generally depend
on Borel parameters in the denominator $1/M_{1,2}$, therefore, for
effectively suppressing the contributions of the excited states
and the higher dimensional operators in OPE, the Borel parameters
$M_{1,2}$ should be neither too large, nor too small. We find the
optimal stability with the requirements shown in Table \ref{T1}.
The regions of Borel parameters which satisfies the requirements
of Table \ref{T1} are shown in Fig.\ref{region} in two-dimensional
diagram of $M_1^2$ and $M_2^2$. We find good stability of the form
factors within these regions.
\begin{tiny}
\begin{table}[h]
\caption{Requirements to select Borel Parameters $M_1^2$ and $M_2^2$
 for each form factors}
\begin{center}
\begin{tabular}{|c|c|c|c|}\hline
Form Factors & contribution  & excited states& excited states  \\
             &of condensate &   of $H$ channel & of $K_0^*(1430)$ channel\\
             \hline
$F_+^{B_sK_0^*}(0)$  & $\le 42\% $ & $\le 10\%$ & $\le 42\%$ \\
\hline $F_+^{DK_0^*}(0)$& $\le 30\% $ & $\le 17\%$ & $\le 24\%$
\\ \hline $F_+^{D_sK_0^*}(0)$ &$\le 28\% $ & $\le 12\%$ & $\le 14\%$  \\ \hline
\end{tabular}\end{center}
\label{T1}
\end{table}
\end{tiny}

In general the higher the dimension of the operators, the smaller
the relevant contributions of the condensates. We find that, in
the numerical analysis, the main contributions to the form factors
are from perturbative term and condensate of dimension-3 operator,
the contributions of operator of dimension 6 are negligible.

The final results for the form factors at $q^2=0$ are
\begin{eqnarray}
F_+^{B_sK_0^*}(0)&=& 0.24\pm 0.10, \nonumber\\
F_+^{DK_0^*}(0)&=&0.57\pm 0.19,\\
F_+^{D_sK_0^*}(0)&=& 0.51\pm 0.20.\nonumber
\end{eqnarray}
The error bars are estimated by the variation of Borel parameters,
the variation of the threshold parameters $s_{1,2}^0$, the
uncertainty of the condensate parameters, the variation of the
quark masses and meson decay constants, and the possible
$\alpha_s$ corrections to the three-point function. The main
contribution comes from the uncertainties of the threshold
parameters and decay constants, which is about $10\%\sim 20\%$ of
the central value, the other uncertainties are only a few percent.
The $\alpha_s$ corrections to two-point functions of $f_B$, $f_D$
and $f_{K_0^*}$ are 10\% to 30\%, we estimate the $\alpha_s$
corrections to three-point function shall be the same order. We
take 30\% as the error caused by the $\alpha_s$ correction in this
work.

Numerically, $F_+^{DK_0^*}$ and $F_+^{D_sK_0^*}$ are slightly
different. The difference is completely caused by the SU(3)
symmetry breaking effect. The numerical result shows that this
effect is about 10\%.

There have been studies of heavy flavor meson to $K_0^*(1430)$
transition form factors in the literature. In \cite{cheng1},
$F_+^{DK_0^*}(0)$ is extracted from the data of the hadronic $D$
decay in the generalized factorization model,
$F_+^{DK_0^*}(0)=1.20\pm 0.07$. It is also calculated in covariant
light-front approach, $F_+^{DK_0^*}(0)=0.48$ \cite{LightFront}.
Compared with these values, our present result
$F_+^{DK_0^*}(0)=0.57\pm 0.19$ is consistent with the light-front
approach.

\begin{figure}[hbt]
\begin{center}
 \epsfig{file=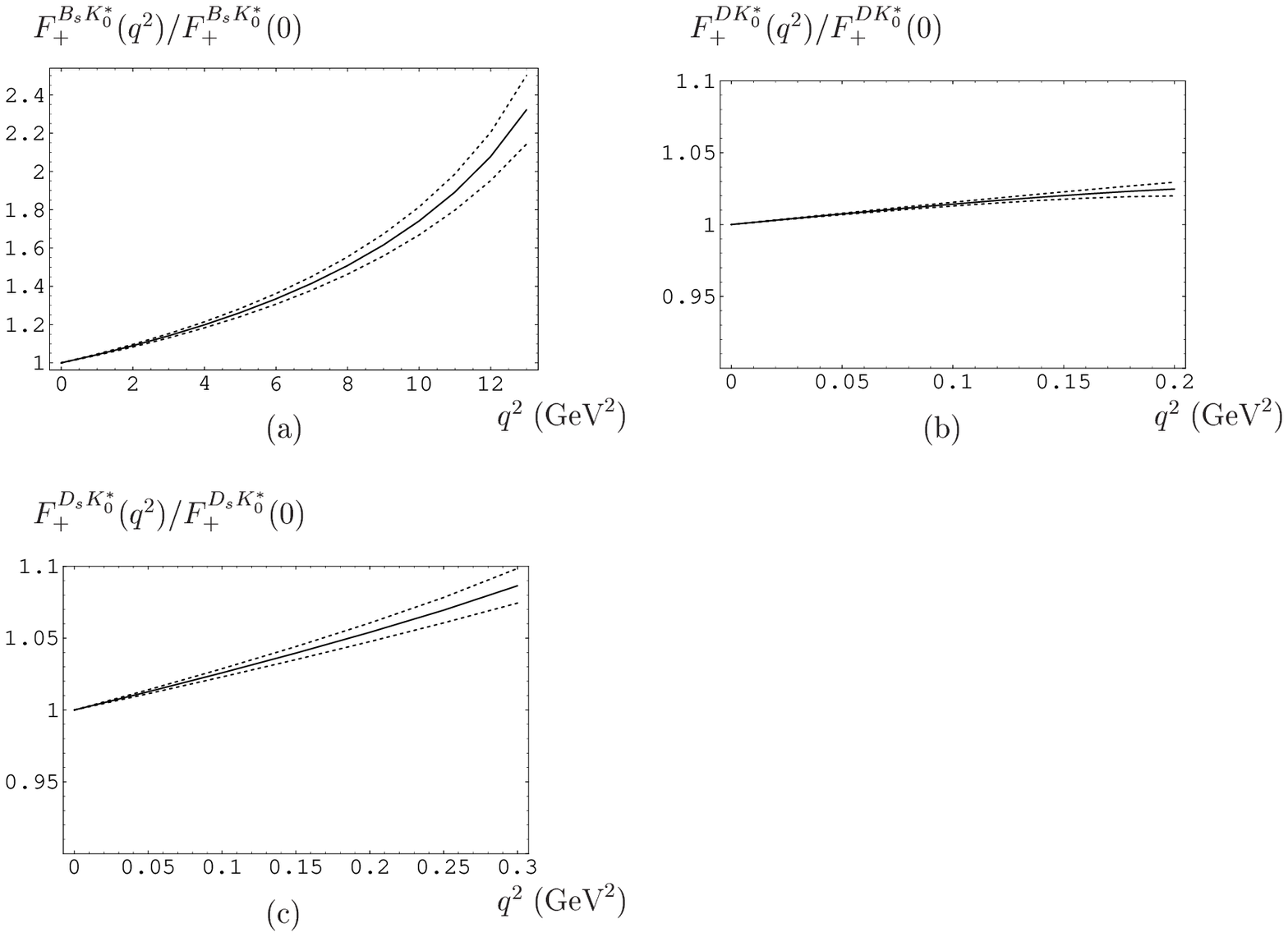, width=14cm,height=10cm}
 \end{center}
\caption{\small $q^2$ dependence of the form factors from QCD sum
rule. The solid curve is for the central value, the dotted curve
is for the errors allowed by the input parameters.} \label{q2}
\end{figure}

The physical region for $q^2$ in $B$ or $D\to
K_0^*(1430)\bar{\ell}\nu$ decay extends from $0$ to
$(m_{B,D}-m_{K_0^*} )^2$, which is not large because of the large
mass of $K_0^*(1430)$. The $q^2$ dependence of the form factors in
the physical range can be calculated directly from QCD sum rule,
which has been discussed in detail in \cite{BBD}. Within the
physical range of $q^2$ in $B$ or $D$ to $K_0^*(1430)$ decays,
there is no non-Landau-type singularity \cite{BBD} with the
threshold parameters $s_1^0$ and $s_2^0$ considered in this paper.
From the sum rule for the form factors listed in the appendix, we
see that the formulas are not singular at $q^2=0$, and $q^2$
appears in the numerator, therefore, if $|q^2|$ is too large, OPE
will fail. So the condition for the OPE be effective is to keep
$|q^2|$ in a moderate range. For the whole physical range of $q^2$
considered in this paper, within the selected window of the Borel
parameters, the contribution of higher-dimensional operators is
effectively small, therefore the OPE is effective.

The form factors as functions of $q^2$, normalized by the value at
$q^2=0~\mbox{GeV}^2$, $F_+(q^2)/F_+(0)$'s are shown in
Fig.\ref{q2}. Within the physical range, the behavior of the form
factor $F_+(q^2)$ is compatible with the pole-model,
$$ F_+(q^2)=\frac{F_+(0)}{1-q^2/m_{\mbox{pole}}^2}. $$

The result of the form factors from QCD sum rule can be fitted
with the pole model. The fitted pole masses are,
\begin{eqnarray}
m^{B_sK_0^*}_{\mbox{pole}}~=& 4.80\pm 0.12~\mbox{GeV},\nonumber\\
m^{DK_0^*}_{\mbox{pole}}~=& 2.9\pm 0.3~\mbox{GeV},\\
m^{D_sK_0^*}_{\mbox{pole}}~=& 1.96\pm 0.12~\mbox{GeV}.\nonumber
\end{eqnarray}

Next we shall use the form factors calculated in QCD sum rule to
study the differential and total decay rates of $B$ and $D\to
K_0^*(1430)\bar{\ell}\nu$ decays. The differential decay rate is
calculated to be
\begin{equation}
\frac{d\Gamma_L}{d
q^2}=\displaystyle\frac{G_F^2|V_{Qq'}|^2}{192\pi ^3
   m_{H}^3}F_+(q^2)^2[
   (m_{H}^2+m_{K_0^*}^2-q^2)^2-4m_H^2m_{K_0^*}^2]^{3/2}.
\end{equation}
The values of the CKM matrix elements are taken in Wolfenstain
parameterization, $A=0.8$, $\lambda =0.22$, $\rho=0.20$, and $\eta
=0.34$ \cite{PDG}, which are relevant to $V_{cs}=0.976$,
$V_{cd}=-0.22$.

\begin{figure}[h]
\begin{center}
 \epsfig{file=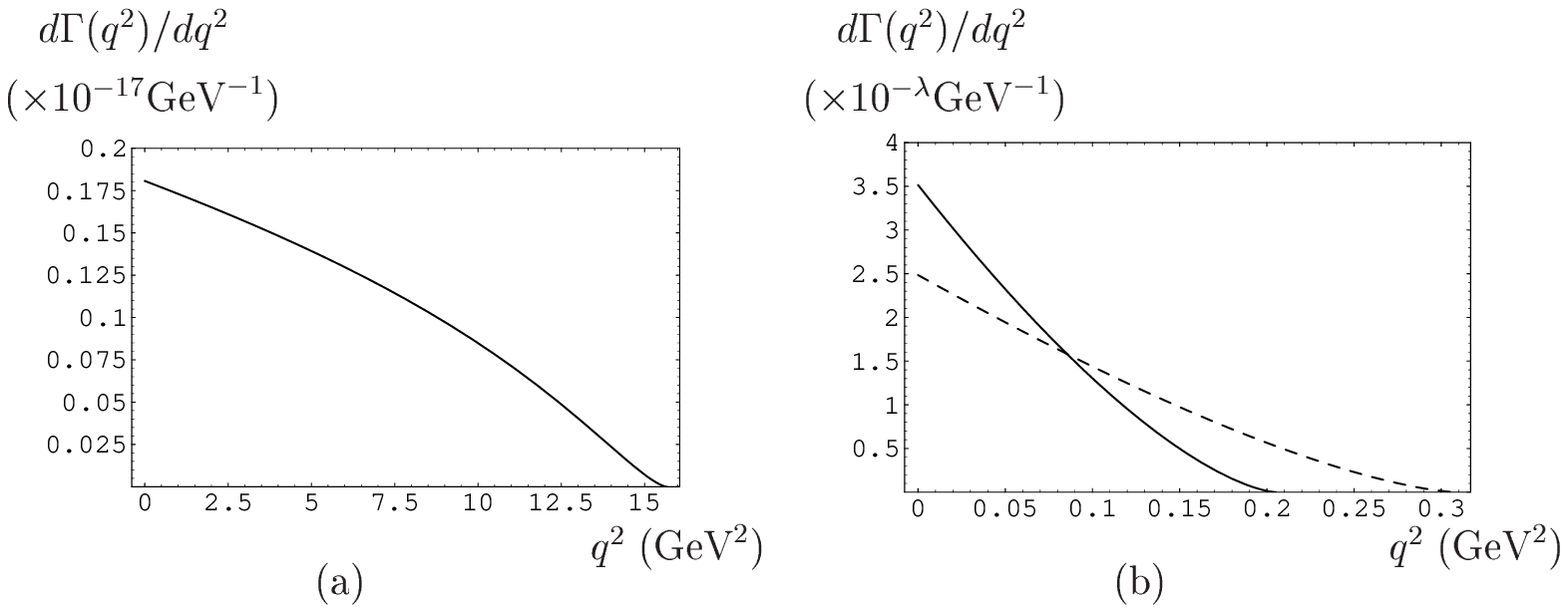, width=14cm,height=5cm}
 \end{center}
\caption{\small Differential decay widths of $B$ and $D$ to
$K_0^*(1430)\bar{\ell}\nu$ as function of $q^2$. (a) For $B_s^0\to
K_0^*(1430)^-\bar{\ell}\nu$; (b) The solid curve is for $D^0\to
K_0^*(1430)^-\bar{\ell}\nu$ and $D^+\to
\bar{K}_0^*(1430)^0\bar{\ell}\nu$ with the scale parameter
$\lambda =15$, while the dashed curve is for $D_s^+\to
K_0^*(1430)^0\bar{\ell}\nu$ with the scale parameter $\lambda
=16$.} \label{gamma}
\end{figure}

The differential decay widths as a function of momentum transfer
squared $q^2$ are shown in Fig.\ref{gamma}. The decay rate of
$D_s^+\to K_0^*(1430)^0\bar{\ell}\nu$ is one more order smaller
than that of $D^0\to K_0^*(1430)^-\bar{\ell}\nu$ and $D^+\to
\bar{K}_0^*(1430)^0\bar{\ell}\nu$ because of the CKM matrix
element suppression of $V_{cd}$ over $V_{cs}$.  After integration
over $q^2$ in the whole physical region, we get the integrated
decay widths
\begin{eqnarray}
&\Gamma (B_s^0\to
K_0^*(1430)^-\bar{\ell}\nu)=(1.6^{+1.7}_{-1.1})\times
10^{-17}~\mbox{GeV},\nonumber\\
&\Gamma (D^{0,+}\to
K_0^*(1430)\bar{\ell}\nu)=(2.9^{+2.3}_{-1.6})\times
10^{-16}~\mbox{GeV},\\
&\Gamma (D_s^+\to
K_0^*(1430)^0\bar{\ell}\nu)=(3.2^{+3.0}_{-2.0})\times
10^{-17}~\mbox{GeV}.\nonumber
\end{eqnarray}
The branching ratio is defined by
\begin{equation}
Br=\Gamma /\Gamma_{total},
\end{equation}
where $\Gamma$ denotes the decay width of each decay mode, and
$\Gamma_{total}$ the total decay width of $B$ or $D$ meson. The
total decay width of one meson is related to its mean life time
$\tau$ by $\Gamma_{total}=\hbar /\tau$. The mean life times of $B$
and $D$ mesons are: $\tau_{B_s}=1.46\times 10^{-12}~s$,
$\tau_{D^{\pm}}=1.04\times 10^{-12}~s$, $\tau_{D^0}=0.41\times
10^{-12}~s$, and $\tau_{D_s}=0.49\times 10^{-12}~s$ \cite{PDG}. We
can obtain the branching ratios
\begin{eqnarray}
&Br(B_s^0\to K_0^*(1430)^-\bar{\ell}\nu)=(3.6^{+3.8}_{-2.4})\times
10^{-5},\nonumber\\
&Br(D^{0}\to K_0^*(1430)^-\bar{\ell}\nu)=(1.8^{+1.5}_{-1.0})\times
10^{-4},\nonumber\\
&Br(D^{+}\to
\bar{K}_0^*(1430)^0\bar{\ell}\nu)=(4.6^{+3.7}_{-2.6})\times
10^{-4},\\
&Br(D_s^+\to K_0^*(1430)^0\bar{\ell}\nu)=(2.4^{+2.2}_{-1.5})\times
10^{-5}.\nonumber
\end{eqnarray}
Although $\Gamma(D^{0}\to
K_0^*(1430)^-\bar{\ell}\nu)=\Gamma(D^{+}\to
\bar{K}_0^*(1430)^0\bar{\ell}\nu)$ under the isospin symmetry, the
branching ratios of these two decay modes are greatly different.
This is due to the large difference of the total decay widths of
$D^0$ and $D^{\pm}$ mesons.

The numerical result shows that the branching ratios of $B$ and
$D$ to $K_0^*(1430)\bar{\ell}\nu$ are rare. Measuring them in
experiment needs high statistics, but these measurements will give
valuable information on the nature of the light scalar mesons.
Recently, Focus collaboration gives an upper limit for the ratio,
$\frac{\Gamma(D^+\to \bar{K}_0^*(1430)^0\mu^+\nu)}{\Gamma (D^+\to
K^-\pi^+\mu^+\nu)}<0.64\%$, in analyzing the mass spectrum of
$D^+\to K^-\pi^+\mu^+\nu$ decay \cite{focus}. By considering the
branching ratio of $Br(D^+\to K^-\pi^+\mu^+\nu)=4.00\pm 0.32 \%$
given in PDG \cite{PDG}, one can get the branching ratio
$Br(D^+\to \bar{K}_0^*(1430)^0\mu^+\nu)<2.8\times 10^{-4}$.
Compared with this upper limit, our result $Br(D^{+}\to
\bar{K}_0^*(1430)^0\bar{\ell}\nu)=(4.6^{+3.7}_{-2.6})\times
10^{-4}$ is compatible with it.  Direct experimental measurement
of these semileptonic branching ratios and more precise
theoretical prediction are highly desired.

 In summary, we have calculated the $B$
and $D$ to $K_0^*(1430)$ transition form factors
$F_+^{B_sK_0^*}(q^2)$, $F_+^{DK_0^*}(q^2)$, $F_+^{D_sK_0^*}(q^2)$
in QCD sum rule, and studied the semileptonic decays of $B_s^0\to
K_0^*(1430)^-\bar{\ell}\nu$, $D^0\to K_0^*(1430)^-\bar{\ell}\nu$,
$D^+\to \bar{K}_0^*(1430)^0\bar{\ell}\nu$, and $D_s^+\to
K_0^*(1430)^0\bar{\ell}\nu$. Measuring these decay modes with high
statistics will give valuable information on the structure of
light scalar mesons.

\vspace{1cm}

{\bf Acknowledgements} This work is supported in part by the
National Science Foundation of China under contract No.10205017,
10575108, and by the Grant of BEPC National Laboratory.

\vspace{1cm}

\begin{center}{\bf Appendix}\end{center}

The analytical results of the coefficients $\hat{B}f_+$ for the
form factors are given here. Two cases are classified, one is for
$B_s$ and $D_s\to K_0^*(1430)$ transitions, the other for $B$ and
$D$ transitions.

{\bf (1)} For $B_s$ and $D_s\to K_0^*(1430)$ transitions, the form
factor in eq.(\ref{formfactor}) will be
\begin{equation}\label{}
F_+(q^2)^{H_sK_0^*}=\frac{m_Q+m_{s}}{f_{K_0^*}m_{K_0^*}m_{H_s}^2f_{H_s}}M_1^2M_2^2
   e^{m_{H_s}^2/M_1^2}e^{m_{K_0^*}^2/M_2^2}\hat{B}f_+^{H_sK_0^*},
  \end{equation}
where $H_s$ denotes $B_s$ or $D_s$, and
\begin{equation}
\hat{B}f_+^{H_sK_0^*}=\hat{B}f_{+H_s}^{pert}+\hat{B}f_{+H_s}^{(3)}+\hat{B}f_{+H_s}^{(5)}
           +\hat{B}f_{+H_s}^{(6)}.
\end{equation}

The perturbative contribution is
\begin{eqnarray}
\hat{B}f_{+H_s}^{pert}&=&\int_{m_s^2}^{s_2^0}ds_2\int_{s_1^L}^{s_1^0}ds_1
  \frac{3}{8M_1^2M_2^2\pi^2\lambda^{3/2}}e^{-\frac{s_1}{M_1^2}-\frac{s_2}{M_2^2}}
  \{q^2(q^2-s_1-s_2)m_Qm_s\nonumber\\
  &&-(q^2-s_1+s_2)m_Q^3m_s+m_Q^2[(q^2+s_1-s_2)s_2+(q^2-s_1+s_2)m_s^2]\nonumber\\
  &&+2q^2[-s_1s_2+(-q^2+s_1+s_2)m_s^2]\},
\end{eqnarray}
where the lower integration limit of $ds_1$ is determined by the
condition that all internal quarks in the perturbative diagram be
on mass shell \cite{Landau},
$$ s_1^L=\frac{m_Q^2}{m_Q^2-q^2}s_2+m_Q^2,$$
and $\lambda=(s_1+s_2-q^2)^2-4s_1s_2$.

The condensate contributions are
\begin{eqnarray}
\hat{B}f_{+H_s}^{(3)}&=&-\frac{\langle
\bar{s}s\rangle}{4M_1^6M_2^6}
 e^{-\frac{m_Q^2}{M_1^2}}\{2M_1^4M_2^4m_Q-M_1^2M_2^4(2M_1^2+m_Q^2)m_s\nonumber\\
 &&+M_2^2m_Q[-M_1^2q^2+(M_1^2+M_2^2)m_Q^2]m_s^2\},
\end{eqnarray}
\begin{eqnarray}
\hat{B}f_{+H_s}^{(5)}&=&-\frac{g\langle \bar{\Psi}\sigma
TG\Psi\rangle}{24M_1^8M_2^4}
 e^{-\frac{m_Q^2}{M_1^2}}\{3M_1^2m_Q[M_1^2(M_1^2+2M_2^2+q^2)-(M_1^2+M_2^2)m_Q^2]\nonumber\\
 &&+(4M_1^2+m_Q^2)[-M_1^2q^2+(M_1^2+M_2^2)m_Q^2]m_s\},
\end{eqnarray}
\begin{eqnarray}
\hat{B}f_{+H_s}^{(6)}&=&-\frac{4\pi\alpha_s\langle\bar{\Psi}\Psi\rangle
^2} {324M_1^8M_2^8(m_Q^2-q^2)} e^{-\frac{m_Q^2}{M_1^2}}
\{M_2^4\{2M_1^4[M_1^2(72M_2^2-17q^2)-q^2(M_2^2\nonumber\\
&&+6q^2)]+M_1^2[-38M_1^4+q^2(24M_2^2+5q^2)+2M_1^2(M_2^2+19q^2)]m_Q^2
\nonumber\\
&&-2[13M_1^4-2M_2^2q^2+M_1^2(12M_2^2+5q^2)]m_Q^4+(5M_1^2-4M_2^2)m_Q^6\}
\nonumber\\
&&+36M_1^6M_2^2m_Q(3M_2^2+q^2-m_Q^2)m_s-18M_1^6(4M_2^4-M_2^2m_Q^2)m_s^2\},
\end{eqnarray}

 {\bf (2)} For $B$ and $D\to K_0^*(1430)$ transitions, the form
factor in eq.(\ref{formfactor}) will be
\begin{equation}\label{}
F_+(q^2)^{H_qK_0^*}=\frac{m_Q}{f_{K_0^*}m_{K_0^*}m_{H_q}^2f_{H_q}}M_1^2M_2^2
   e^{m_{H_q}^2/M_1^2}e^{m_{K_0^*}^2/M_2^2}\hat{B}f_+^{H_qK_0^*},
  \end{equation}
where $H_q$ denotes $B$ or $D$, and
\begin{equation}
\hat{B}f_+^{H_qK_0^*}=\hat{B}f_{+H_q}^{pert}+\hat{B}f_{+H_q}^{(3)}+\hat{B}f_{+H_q}^{(5)}
           +\hat{B}f_{+H_q}^{(6)}.
\end{equation}

The perturbative contribution is
\begin{eqnarray}
\hat{B}f_{+H_q}^{pert}&=&\int_{m_s^2}^{s_2^0}ds_2\int_{s_1^L}^{s_1^0}ds_1
  \frac{-3}{8M_1^2M_2^2\pi^2\lambda^{3/2}}e^{-\frac{s_1}{M_1^2}-\frac{s_2}{M_2^2}}
  \{s_2(-q^2-s_1+s_2)m_Q^2\nonumber\\
  &&+[(s_1-s_2)^2-q^2(s_1+s_2)]m_Qm_s+(q^2-s_1+s_2)m_Q^3m_s\nonumber\\
  &&+s_1[2q^2s_2-(q^2-s_1+s_2)m_s^2]\},
\end{eqnarray}

The condensate contributions are
\begin{eqnarray}
\hat{B}f_{+H_q}^{(3)}&=&-\frac{\langle
\bar{q}q\rangle}{2M_1^2M_2^2}
 e^{-\frac{m_Q^2}{M_1^2}-\frac{m_s^2}{M_2^2}}(m_Q-m_s),
\end{eqnarray}
\begin{eqnarray}
\hat{B}f_{+H_q}^{(5)}&=&-\frac{g\langle \bar{\Psi}\sigma
TG\Psi\rangle}{8M_1^6M_2^6}
 e^{-\frac{m_Q^2}{M_1^2}-\frac{m_s^2}{M_2^2}}\{-M_2^2(M_1^2+M_2^2)m_Q^3
 -M_1^2M_2^2(2M_1^2+M_2^2\nonumber\\
&&+q^2)m_s+M_2^2(M_1^2+M_2^2)m_Q^2m_s
 +M_1^2m_Q[M_2^2(M_1^2+2M_2^2+q^2)\nonumber\\
&&-(M_1^2+M_2^2)m_s^2]\}
\end{eqnarray}
\begin{eqnarray}
\hat{B}f_{+H_q}^{(6)}&=&-\frac{4\pi\alpha_s\langle\bar{\Psi}\Psi\rangle
^2} {324M_1^8M_2^6(m_Q^2-q^2)} e^{-\frac{m_Q^2}{M_1^2}}
\{2M_1^2[2M_2^2(11M_2^2-q^2)q^2-M_1^2q^2(6M_2^2+q^2)\nonumber\\
&&+M_1^4(54M_2^2+23q^2)]m_Qm_s-2[23M_1^6+7M_2^4q^2-2M_1^4(3M_2^2+q^2)\nonumber\\
&&+M_1^2(22M_2^4-5M_2^2q^2)]m_Q^3m_s-2(M_1^4+3M_1^2M_2^2-7M_2^4)m_Q^5m_s\nonumber\\
&&+(5M_1^2-4M_2^2)m_Q^6(m_2^2-m_s^2)-M_1^2m_Q^2\{-M_2^2[-38M_1^4+q^2(24M_2^2+5q^2)\nonumber\\
&&+2M_1^2(M_2^2+19q^2)]+[58M_1^4+q^2(27M_2^2+5q^2)+6M_1^2(2M_2^2+7q^2)]m_s^2\}\nonumber\\
&&+m_Q^4\{-2M_2^2[13M_1^4-2M_2^2q^2+M_1^2(12M_2^2+5q^2)]+[29M_1^4-4M_2^2q^2\nonumber\\
&&+m_1^2(27M_2^2+10q^2)]m_s^2\}+M_1^4\{2M_2^2[M_1^2(72M_2^2-17q^2)-q^2(M_2^2+6q^2)]\nonumber\\
&&+[-4m_1^2(18M_2^2-19q^2)+q^2(12M_2^2+13q^2)]m_s^2\}\} ,
\end{eqnarray}

\end{document}